\documentclass{aastex63}
\usepackage{color} 
\usepackage{CJK}

\newcommand{\target}{KIC 4544587}

\received{mm dd, yyyy}
\revised{mm dd, yyyy}
\accepted{mm dd, yyyy}
\submitjournal{ApJ}

\shorttitle{contribution of dynamic tides to apsidal motion}
\shortauthors{Ou et al.}

\begin{document}
\begin{CJK}{UTF8}{gbsn}

\title{The Measurement of Dynamic Tidal Contribution to Apsidal Motion in Heartbeat Star KIC 4544587}

\correspondingauthor{Jian-wen Ou, Cong Yu}
\email{oujw3@mail.sysu.edu.cn, yucong@mail.sysu.edu.cn}

\author[0000-0002-6176-7745]{Jian-wen Ou}
\affil{School of Physics and Astronomy, Sun Yat-Sen University, Zhuhai, 519082, China}
\affil{Key Laboratory of Modern Astronomy and Astrophysics in Ministry of Education, Nanjing University, Nanjing, 210093, China}

\author[0000-0003-0454-7890]{Cong Yu}
\affil{School of Physics and Astronomy, Sun Yat-Sen University, Zhuhai, 519082, China}

\author[0000-0002-6926-2872]{Ming Yang}
\affil{Key Laboratory of Modern Astronomy and Astrophysics in Ministry of Education, Nanjing University, Nanjing, 210093, China}
\affil{School of Astronomy and Space Science, Nanjing University, Nanjing, 210093, China}

\author[0000-0002-7614-1665]{Chen Jiang}
\affil{Max-Planck-Institut f\"ur Sonnensystemforschung, G\"ottingen, 37077 , Germany}

\author{Bo Ma}
\affil{School of Physics and Astronomy, Sun Yat-Sen University, Zhuhai, 519082, China}

\author{Guan-fu Liu}
\affil{School of Physics and Astronomy, Sun Yat-Sen University, Zhuhai, 519082, China}

\author{Shang-fei Liu}
\affil{School of Physics and Astronomy, Sun Yat-Sen University, Zhuhai, 519082, China}

\author{Juan-juan Luo}
\affil{School of Physics and Electronics, Qiannan Normal University for Nationalities, Duyun, 558000, China}

\begin{abstract}

Apsidal motion is a gradual shift in the position of periastron. The impact of dynamic tides on apsidal motion has long been debated, because the contribution could not be quantified due to the lack of high quality observations. KIC 4544587 with tidally excited oscillations has been observed by \textit{Kepler} high-precision photometric data based on long time baseline and short-cadence schema. In this paper, we compute the rate of apsidal motion that arises from the dynamic tides as $19.05\pm 1.70$ mrad yr$^{-1}$ via tracking the orbital phase shifts of tidally excited oscillations. We also calculate the procession rate of the orbit due to the Newtonian  and general relativistic contribution as $21.49 \pm 2.8$ and $2.4 \pm 0.06$ mrad yr$^{-1}$, respectively. The sum of these three factors is in excellent agreement with the total observational rate of apsidal motion $42.97 \pm 0.18$ mrad yr$^{-1}$ measured by eclipse timing variations. The tidal effect accounts for about 44\% of the overall observed apsidal motion and is comparable to that of the Newtonian term. Dynamic tides have a significant contribution to the apsidal motion. The analysis method mentioned in this paper presents an alternative approach to measuring the contribution of the dynamic tides quantitatively.

\end{abstract}


\section{Introduction} \label{sec1}

In close binary systems with eccentric orbits, the tidal and rotational distortions of the components induce mass redistributions in the star from spherical symmetry. The non-spherical gravitational field leads to a gradual shift in the position of periastron ($\omega$), known as apsidal motion \citep[$\dot{\omega}$, ][]{cla1993,zas2020}. The analogue to apsidal motion is the well-known effect as Mercury's perihelion advance in the solar system.

Theoretically two different contributions are thought to devote additively to such precession: a classical or Newtonian term \citep[$\dot{\omega}_{\rm NT}$,][]{ste1939} arising from the distributions of density within components, and a general relativistic correction \citep[$\dot{\omega}_{\rm GR}$,][]{sha1985} which are on account of the space distortion in a strong gravitational field. Near the periastron, the tidal distortion leads to a perturbation of the external gravitational field, which in turn contributes to the apsidal motion and is called the contribution of the dynamic tides \citep[$\dot{\omega}_{\rm DT}$,][]{sme1991}. The Newtonian term and the general relativistic term have been widely confirmed \citep{spl2002,con2013, man2020}. Unfortunately, the scenario of dynamic tides was not always fulfilled \citep{cla2002,sem2005,pav2011,cla2019} because tidal contribution could not be quantified due to the lack of high quality observations.

Heartbeat stars are close eccentric binary stars in short orbital period with tidal distortions \citep{tho2012}. Most of the heartbeat stars undergo periodic tidal force throughout the orbit, which cause dynamic tides and induce tidally excited oscillations (TEOs) simultaneously \citep{shp2016, ful2017}. Tidal effects are most significant at the position of periastron causing a prominent distorting of the shape of the light curve, known as ``heartbeat" signature.

\textit{Kepler} space telescope \citep{bor2010} provides us long time baseline and high-precision photometric data that are ideal for the study of the dynamic tides in heartbeat stars \citep{dim2017, cheng2020}, especially the short-cadence (58.89 seconds) data. Over 170 heartbeat stars have been discovered in \textit{Kepler} data \citep{kir2016, gau2019}, and some targets have been analyzed in detail, such as KOI-54 the prototypical heartbeat star \citep{wel2011, ole2014}, KIC 8164262 \citep{ham2018}, KOI-3890 \citep{kus2019}, KIC 4142768 \citep{guo2019}, and KIC 5006817 \citep{mer2021}.

More recently, an excellent study for the relationship between the position of periastron and the pulsation phase of TEOs has been discussed in \cite{guo2020}. However, the  position of periastron that they discussed was a fixed value, that is, $\omega$ did not change in eccentric orbits. Here, we focus on the precession of periastron $\dot{\omega}$, and investigates the dynamic tidal contribution to the rate of apsidal motion quantitatively. The changes in the position of periastron cause the changes in the induced time of TEOs. Therefore, the measurement of the time delay of TEOs in heartbeat stars makes it possible to probe the relationship between the tidal effect and the apsidal motion.

KIC 4544587 is an eccentric close binary system with dynamical tides that presents TEOs. On account of the nature of stellar oscillations and apsidal motion, we deduce the rate of apsidal motion  arising from dynamic tides. This paper is organized as follows. In Section \ref{sec2} we describe the observations of KIC 4544587 including \textit{Kepler} photometry data and the absolute parameters of this system. In Section \ref{sec3}, the short-cadence photometry data  is used for light curve analysis. Measuring the primary and secondary mid-eclipse times is described in Section \ref{sec3.1}, and the principle of TEOs' phase delay is presented in Section \ref{sec3.2}. Based on the measurement of eclipse timing variations, we calculated the total apsidal motion rate of KIC 4544587 in Section \ref{sec4.1}. Newtonian term and general relativistic contributions relied on accurate absolute parameters are calculated in Section \ref{sec4.2}. Finally, we use the shifts of dynamic tide to derive the contribution of apsidal motion caused by tidal effects in Section \ref{sec4.3}. The main results and discussions are summarized in Section \ref{sec5}.

\section{The observations of \target} \label{sec2}
\subsection{\textit{Kepler} photometry} \label{sec2.1}
\begin{figure}[htb]
\center
  \includegraphics[width=0.68\textwidth]{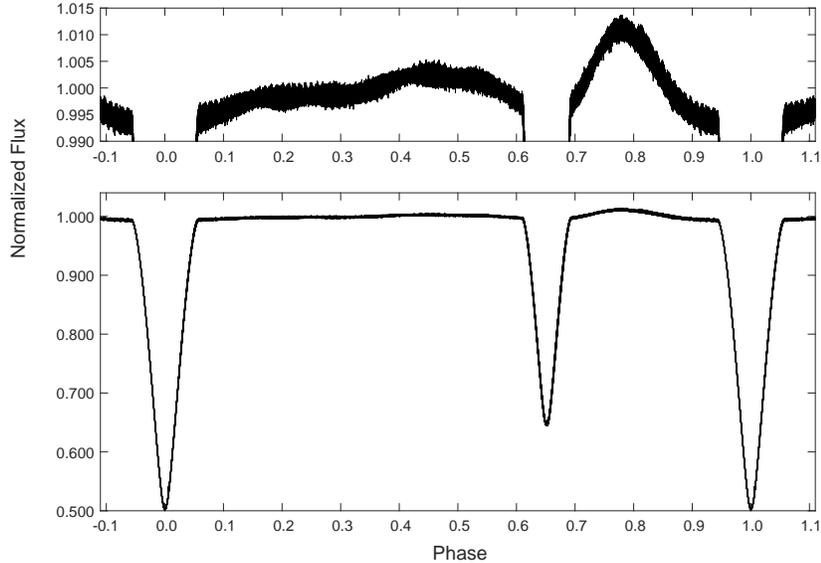}
  \caption{The detrended light curve for the short-cadence data of quarters 7-10. \textit{Top} panel is a magnified image of the out-of-eclipse part that shows a remarkable heartbeat signature at the phase of approximately $0.78$, and tidally excited oscillations throughout the orbit due to the periodic tidal forces.}
  \label{Fig1}
\end{figure}

KIC 4544587 has a \textit{Kepler} magnitude $K_p = 10.8$. The photometric data include long-cadence  during quarters 0-17 and short-cadence during quarters 3.2, and 7-10. The short-cadence data with 58.89 seconds sampling has the advantage of increasing frequency resolution. High-frequency resolution enables the accurate identification of tidally excited oscillations which present in this target. Consequently, short-cadence data during quarters 7-10 is used to analyze the TEOs and eclipse timing variations. As a preprocessing, we detrended and normalized these data by fitting a first- or second-order polynomial to individual segments of data separated by gaps, which caused by spacecraft rolls or safe mode events. Figure \ref{Fig1} shows the orbital phase of the detrended light curve and its magnified image.

\subsection{The absolute dimensions} \label{sec2.2}

Accurate absolute dimensions are necessary for reliable interpretation of the apsidal motion. Thanks to the excellent work of \cite{ham2013}, we can easily retrieve the basic information related to KIC 4544587. The atmospheric parameters for the components, such as the effective temperatures and surface gravities, are derived from the disentangled spectra data. The projected rotational velocities are determined by the optimal fitting of the metal lines. Spectral information shows that the components of the binary system are all A-type stars. The orbital period and time of primary minimum are determined with the short-cadence data (quarters 3.2, 7 and 8) using KEPHEM software \citep{prs2011}, and the geometrical configuration and stellar parameters are obtained by fitting the light curve for quarters 7-8 with PHOEBE package \citep{prs2005}. We summarize the parameters in Table \ref{tabxxx1}.

The results of the pulsational analysis are also presented in \cite{ham2013}. KIC 4544587 includes at least one pulsating component. Asteroseismology analysis revealed 31 pulsating modes, including 23 self-excited pressure and gravity modes, and 8 tidally excited modes that are multiples of the orbital frequency. The maximum TEOs frequency is 44.31 d$^{-1}$ that corresponds to 97 times of the orbital frequency. A first look by \cite{ham2013} suggested that the rapid apsidal motion 43.06 mrad yr$^{-1}$ may be partially attributable to the resonant oscillations.

\begin{table}
  \caption{The observational parameters of heartbeat star KIC 4544587 taken from \cite{ham2013}.}
  \label{tabxxx1}
  \begin{center}
  \begin{tabular}{ccc}
  \hline
 & Primary & Secondary\\
\hline
Mass ($M_\odot$) & $1.98\pm 0.07$ & $1.60 \pm 0.06$\\
Radius ($R_\odot$) & $1.76\pm 0.03$ & $1.42 \pm 0.02$\\
$T_{\rm eff}$ (K) & $8600\pm 100$ & $7750 \pm 180$\\
log $g$ (cgs) & $4.241\pm 0.009$ & $4.33 \pm 0.01$\\
$v$ sin $i$ (km s$^{-1}$) & $86.5\pm 1.3$ & $75.8 \pm 1.5$\\
\hline
 & \multicolumn{2}{c}{System}  \\
 \hline
 Semimajor axis ($R_\odot$) & \multicolumn{2}{c}{$10.855\pm0.046$} \\
 Orbital period (d) & \multicolumn{2}{c}{$2.189094\pm0.000005$} \\
 Orbital eccentricity & \multicolumn{2}{c}{$0.275\pm0.004$} \\
 Orbital inclination ($^\circ$) & \multicolumn{2}{c}{$87.9\pm0.03$} \\
Argument of periastron (rad) & \multicolumn{2}{c}{$5.74\pm0.03$} \\
Apsidal advance (mrad yr$^{-1}$) & \multicolumn{2}{c}{$43.06\pm0.02$} \\
Time of primary minimum (BJD) & \multicolumn{2}{c}{$2455462.006137\pm0.000009$} \\
  \hline
\end{tabular}
\end{center}
\end{table}

\section{Light curve analysis} \label{sec3}
\subsection{Measuring mid-eclipse times} \label{sec3.1}
\begin{figure}[htb]
\center
  \includegraphics[width=0.5\textwidth]{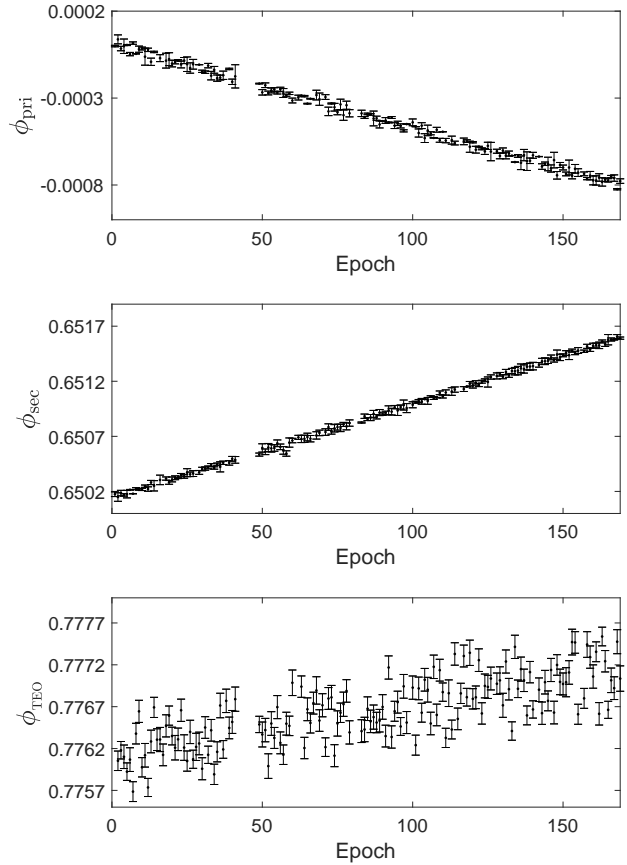}
  \caption{\textit{Top} and \textit{middle} panels are the orbital phase of the primary and the secondary mid-eclipse times (Section \ref{sec3.1}), respectively. The opposing direction of the primary and the secondary mid-eclipse phases demonstrates a classical apsidal motion. \textit{Bottom} panel depicts the shifts of dynamic tides  in units of orbital phase (see details in Section \ref{sec3.2}). }
  \label{Fig2}
\end{figure}

We use a polynomial fit to determine the mid-eclipse times as described in \cite{rap2013} and \cite{yan2015}. The analytic function does not represent a physical model, but rather searches a local minimum flux $F_0$ on the eclipsing binary light curve. The polynomial template defines the stellar flux $F$ as a function of time $t$ by

\begin{equation}
F=\alpha (t-t_0)^2 + \beta (t-t_0)^4 + F_0 \ , \label{equ1}
\end{equation}
where $\alpha$, $\beta$ are the coefficients, and $t_0$ is the mid-eclipse time. By fitting the mean phased shape of all eclipse light curves, we obtain the theoretical template which depends on four parameters $\alpha$, $\beta$, $t_0$ and $F_0$. Next, the whole light curve is divided into many individual eclipse parts for each epoch, i.e. the number of orbits $N$. All the other parameters except $t_0$ are fixed to be the same as the theoretical template. Then, the fitted polynomial parameter $t_0$ is used to determine a more accurate observed mid-eclipse time in each individual eclipses. We take this approaches to find the horizontal shifts of the primary and secondary eclipse, as shown in Figure \ref{Fig2} the \textit{top} and \textit{middle} panel, respectively. The $x$-axis is the orbital epoch, whilst the $y$-axis is the mid-eclipse phase of orbit. The time of primary minimum in Barycentric Julian Days (BJD) chooses as $2455462.006137$ which is consistent with the choice made in \cite{ham2013}, and we set the primary eclipse as orbital phase 0.

\begin{figure}[htb]
\center
  \includegraphics[width=0.7\textwidth]{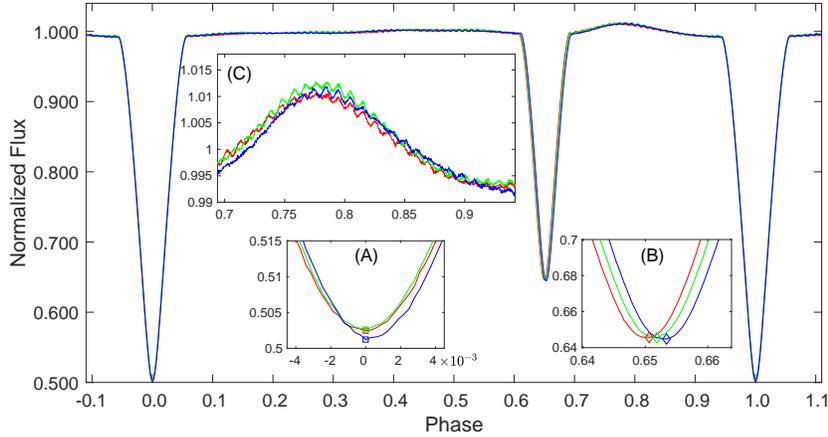}
  \caption{The orbital phase light curve of three epochs. Red, green and blue lines indicate  epochs $N = 7$, 85, and 163, respectively ($\Delta N = 78$). Inset-A, the primary minimum eclipses are fixed at orbital phase 0. Inset-B, the position of the secondary eclipse showing significant horizontal shifts. Inset-C, the horizontal shift of the dynamic tides. The change of spacing between primary and secondary eclipse deduces the total rate of apsidal motion (Equation \ref{equ5}). The horizontal shift of dynamic tides in orbital phase is obtained by tracing the delayed phase of TEOs, which can be used to deduce its contribution to apsidal motion (Equation \ref{equ6}).}
  \label{Fig2add}
\end{figure}

\subsection{Measuring phase delay of TEOs} \label{sec3.2}

The relationship between the pulsation intrinsic phase $\varphi_{\rm I}$ of TEOs and the argument of periastron $\omega$ has been well discussed in \cite{guo2020}(his Equation 3). Here we focus on the induce time delay of the pulsation phase due to the change of periastron position. At the time of periastron passage, the tidal forces induce pulsation modes with intrinsic phase $\varphi_{\rm I}$. When a TEO with frequency $\nu$ dominates the dynamical tidal response, a sinusoidal luminosity fluctuation will be generated as the form \citep{ful2017},  

\begin{equation}
{\Delta L} \propto {\rm sin}[2 \pi \nu (t-t_{\rm p}) + \varphi_{\rm I}] \ ,
\end{equation}
where $t_{\rm p}$ is the passing time of periastron. If the position of periastron has a small displacement, the tidally induced pulsations will be postponed for the time $t'$, and the luminosity fluctuation then becomes,
\begin{eqnarray}
 \Delta L  \propto  {\rm sin}[2 \pi \nu (t-t_{\rm p}+t') + \varphi_{\rm I}] &=& {\rm sin}[2 \pi \nu (t-t_{\rm p}) + \varphi_{\rm I} +  \varphi_{\rm D}] \ ,  \\
  \varphi_{\rm D} & \equiv & 2 \pi \nu t' \ , \label{equ_phiD}
 \end{eqnarray}
with $t^\prime$ being the time of periastron displacement and accordingly $\varphi_{\rm D}$ being the delayed phase of pulsation due to the delay of induce time. In turn, the measurements of $t'$ can be reversely deduced from the delayed phase $\varphi_{\rm D}$ of TEOs.

The delayed phase $\varphi_{\rm D}$ is proportional to the frequency of TEOs. In order to get $\varphi_{\rm D}$  accurately, we choose the maximum TEOs frequency $44.31 {\rm \ d}^{-1}$ to track the pulsation phase of the light curve. The prescription of the phase modulation method \citep{mur2014} is applied on this specific measurement. In practice, we divide the entire light curve into $N$ short segments according to each epoch, and then calculate the frequency of each segment in the Fourier transform. The peak frequency closest to $44.31 {\rm \ d}^{-1}$ is selected. Subsequently, a least-squares fit at the peak frequency is carried out for each segment. The fitting model is a sine function that includes phase as a free parameter. Finally, the measured delayed phase $\varphi_{\rm D}$ and its uncertainty are directly obtained from the fitting procedure. In this analysis, only the out-of-eclipse light curve is considered in order to decrease the contamination of orbital frequency.

We emphasize that the intrinsic phase $\varphi_{\rm I}$ of TEOs is not concerned here. Instead, we focus on the delayed phase $\varphi_{\rm D}$ that reflects the changes of orbital phase. To illustrate the change of the delay phase $\varphi_{\rm D}$ on orbit, we arbitrarily choose the obvious curves and have plotted Figure \ref{Fig2add}. Taking the primary minimum eclipse as the reference point and fixing it at orbital phase 0 (inset-A), the orbital phase of the light curve for epochs $N = 7,\ 85,\ 163$ were plotted by red, green and blue lines, respectively. These lines represent the early, middle and late stages of the orbit. It can be seen from the inset-B that the position of the secondary eclipse has significant horizontal shifts. The total precession rate of apsidal motion is deduced from the spacing between the primary and secondary eclipses, and the details are described in Section \ref{sec4.1}.

The horizontal shift of the dynamic tides is displayed in inset-C. The specific value of dynamic tides shift is obtained by the delayed phase $\varphi_{\rm D}$ of TEOs and then converted to the orbital phase:

\begin{equation}
\phi_{\rm TEO} = \phi_{\rm TEO0} + \Delta \phi_{\rm TEO} = (t_{\rm p} + t')/P_{\rm a} = (t_{\rm p} + \frac{\varphi_{\rm D}}{2\pi \nu})/ P_{\rm a} \ ,
\label{equ_add1}
\end{equation}
where, $P_{\rm a}$ is the anomalistic period \footnote{$P_{\rm a}$ defines the interval of time between two consecutive periastron \citep{gim1995}. Since the time of two consecutive periastron cannot be obtained, here we take $P_{\rm a}$ as the average orbital period of quarters 7-10 data, that is, $P_{\rm a} = P_{\rm orb} = 2.189109 \ {\rm d}$.}, and $\varphi_{\rm D}$ is measured by the phase modulation method. $\phi_{_{\rm TEO0}} = t_{\rm p}/P_{\rm a}$ is the periastron passing time in units of orbital phase. To avoid confusion, we label the pulsation phase as $\varphi$ and the orbital phase as $\phi$, respectively. This approach produced a set of orbital phases $\phi_{\rm TEO}$ as  function of  epoch. The measurements of $\phi_{\rm TEO}$ are shown in the \textit{bottom} panel of Figure \ref{Fig2}.

\section{The argument of periastron}\label{sec4}
\subsection{Determining total rate of apsidal motion by mid-eclipse times}\label{sec4.1}
The eclipse minima of the light curve in KIC 4544587 are narrow and deep enough to permit getting  accurate mid-eclipse times, which can be used to precisely determine the total effect of apsidal motion in this binary system. The method presented by \cite{gim1995} is often used to calculate the apsidal motion rate in eclipsing binary systems. This method assumes that the binary system has a nonzero eccentricity $e$ and the argument of periastron $\omega$ is precessing uniformly. The mid-eclipse time of the primary $t_{\rm pri}$ and the secondary $t_{\rm sec}$ are expressed \citep{gim1995, yee2020}:

\begin{eqnarray}
t_{\rm pri}(N) &=&  t_0 + N P_{\rm s} - \frac{e P_{\rm a}}{\pi} {\rm cos}\ \omega(N) \ , \label{tpri} \\ 
t_{\rm sec}(N) &=&  t_0 + N P_{\rm s} + \frac{e P_{\rm a}}{\pi} {\rm cos}\ \omega(N) + \frac{P_{\rm a}}{2}\ , \label{tsec} \\ 
\omega &=& \omega_0 + N \dot{\omega} \ , \label{w3}
\end{eqnarray}
where $P_{\rm s} = P_{\rm a} (1-{\dot{\omega}}/{2 \pi})$ is the sidereal period, and $\dot{\omega}$ is the rate of apsidal motion in units of rad per cycle. Subtracting Equation (\ref{tpri}) from Equation (\ref{tsec}) generates:

\begin{equation}
t_{\rm sec} - t_{\rm pri} = \frac{2 e P_{\rm a}}{\pi} {\rm cos}\ \omega + \frac{P_{\rm a}}{2}\ . \label{equ_sub}
\end{equation}

With some rearrangements the time interval between primary and secondary mid-eclipses gives the relationship of $e$ and $\omega$ as:

\begin{equation}
e\ {\rm cos}\ \omega=\frac{\pi}{2} \left( \frac{t_{\rm sec}-t_{\rm pri}}{P_{\rm a}}-\frac{1}{2} \right) = \frac{\pi}{2} (\phi_{\rm sec}-\phi_{\rm pri}-0.5)\ . \label{equ5}
\end{equation}

\begin{figure}[htb]
\center
  \includegraphics[width=0.6\textwidth]{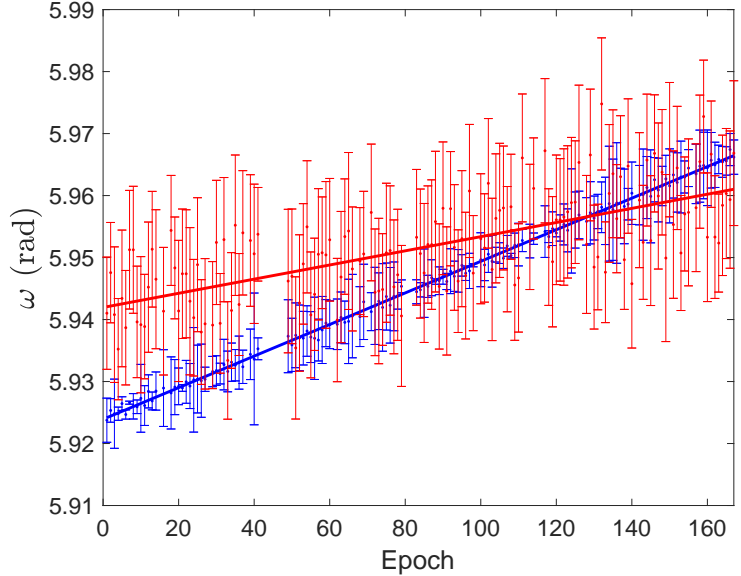}
  \caption{Depicted are the position shift of periastron measured by mid-eclipse times (blue dots) and tidally excited oscillations (red dots), respectively. The slope of periastron arguments caused by dynamic tides is less than that measured by mid-eclipse times, implying that the dynamic tides partially contribute to the total of apsidal motion.}
  \label{Fig3}
\end{figure}

The parameters on the right hand side of Equation (\ref{equ5}) have been measured from the  light curve (Figure \ref{Fig2}). Eccentricity has a major impact on solving the argument of periastron. Since the photometric light curve used in this analysis is different from those of \cite{ham2013}, here we use PHOEBE package \citep{prs2005, con2020} that containing the tidal distortion effects to determine the orbit eccentricity again. The acquired eccentricity $e = 0.252$ is fixed in the following analysis, because the perturbation of eccentricity is in much longer time scale than the dynamical motions considered here. Finally, the observed value of $\omega$ is obtained through the solution of Equation (\ref{equ5}). The results are marked by blue dots as shown in Figure \ref{Fig3}. The increase of $\omega$ is a remarkable manifestation of the apside line precession. By using Equation (\ref{w3}) to fit the values of $\omega$, the total rate of apsidal motion is determined as $\dot{\omega}_{\rm tot} = 42.97 \pm 0.18$ mrad yr$^{-1}$ which is in agreement with the literature value $43.06 \pm 0.02$ mrad yr$^{-1}$ reported by \cite{ham2013}.

\subsection{The contributions of Newtonian term and general relativistic}\label{sec4.2}

In the theoretical framework of apsidal motion, the total rate is divided into two parts: the Newtonian effects and the general relativistic corrections. Supposing the stellar rotation axis is aligned with the normal of the orbital plane, the rate of apsidal motion contributed by the Newtonian term (NT) takes the form of \citep{sha1985,ros2020}:

\begin{equation}
\dot{\omega}_{\rm NT} = \frac{2\pi}{P_{\rm orb}} \left[ 15h(e) \left\{ \frac{k_{2,1}}{q} \left ( \frac{R_1}{a} \right )^5 + k_{2,2}q \left ( \frac{R_2}{a} \right )^5 \right\} + g(e) \left \{ k_{2,1} \frac{1+q}{q} \left ( \frac{R_1}{a} \right )^5  \left ( \frac{P_{\rm orb}}{P_{\rm rot,1}} \right )^2 + k_{2,2}(1+q)\left ( \frac{R_2}{a} \right )^5  \left ( \frac{P_{\rm orb}}{P_{\rm rot,2}} \right )^2  \right \} \right]  \ , \label{equ10} 
\end{equation}

\begin{eqnarray} 
h(e) &=& \frac{1+3e^2/2+e^4/8}{(1-e^2)^5}\ , \label{equ11} \\ 
g(e) &=& \frac{1}{(1-e^2)^2}\ , \label{equ12}
\end{eqnarray}
where, the subscripts ``1" and ``2" refer to the primary and the secondary star, respectively. $R$ is the stellar radius. $P_{\rm rot}$ is the rotational period. $P_{\rm orb}$ is the orbital period, and $q = m_1/m_2$ is the mass ratio. In this paper, we only consider the perturbations arising from the second order harmonic distortions \citep{ste1939} of the gravitational potential, the term of internal structure constant denoted by $k_{2}$.

Using the stellar evolution models computed by \cite{cla2004}\footnote{The retrieved form is available online, https://vizier.u-strasbg.fr/viz-bin/VizieR-3?-source=J/A\%2bA/424/919/tables.} we retrieve the internal structure constants as $k_{2,1} = 4.087\times 10^{-3}$ and $k_{2,2} = 4.078\times 10^{-3}$  for the primary and secondary components, respectively, which correspond to A-type stars \citep{ham2013}. Using the binary parameters of KIC 4544587 along with its uncertainties, we obtain a value of $\dot{\omega}_{\rm NT} = 21.49 \pm 2.80$ mrad yr$^{-1}$ for  Newtonian contribution to the total rate of apsidal motion. It is worthy of note that the estimated value $\dot{\omega}_{\rm NT}$ here is the upper limit because we assume that the stellar rotation axis is parallel to the orientation of the  orbital angular momentum in the first place.

The general relativistic (GR) correction to the total rate of apsidal motion is given by the expression \citep{sha1985,ros2020}:

\begin{equation}
\dot{\omega}_{\rm GR} = \frac{2\pi}{P_{\rm orb}} \frac{3G(m_1+m_2)}{c^2a(1-e^2)} = \left( \frac{2\pi}{P_{\rm orb}} \right) ^ {5/3} \frac{3[G(m_1+m_2)]^{2/3}}{c^2(1-e^2)} \ ,  \label{equ9}
\end{equation}
where $G$ is the gravitational constant, $c$ the speed of light, and $a$ the semi-major axis. Using the observed parameters of the binary system, we figure out $\dot{\omega}_{\rm GR}$ as  $2.40 \pm 0.06$ mrad yr$^{-1}$.  Therefore, the contribution of the sum of the Newtonian effects and the general relativistic corrections is $23.89$ mrad yr$^{-1}$, which is significantly less than the total rate of $42.97$ mrad yr$^{-1}$, indicating that there must be some missing factors influencing the apsidal motion. 

 Firstly, the third light contamination value for KIC 4544587 is estimated to be 0.019. Note that 1 implies complete contamination and 0 implies no contamination of the CCD pixels. This contamination value suggests that KIC 4544587 suffers no impact from the third body. Secondly, the rotation axis of the primary more or less misaligns with the normal of the orbital plane \citep{kha2007, pav2011,sch2016}. If the angle between the rotation axis and the orbit plane is taken into account, the contributions of Newtonian will be less than $21.49\pm2.8$ mrad yr$^{-1}$. Such angle will increase the deviation, so it cannot be caused by misalignments between rotation and orbit axes. Finally, the internal structure constants of each component are consistent with their spectral types. The internal structure constants are unlikely to make such a big deviation.

\subsection{The contribution of dynamic tides to apsidal motion}\label{sec4.3}

\begin{figure}[htb]
\center
  \includegraphics[width=0.5\textwidth]{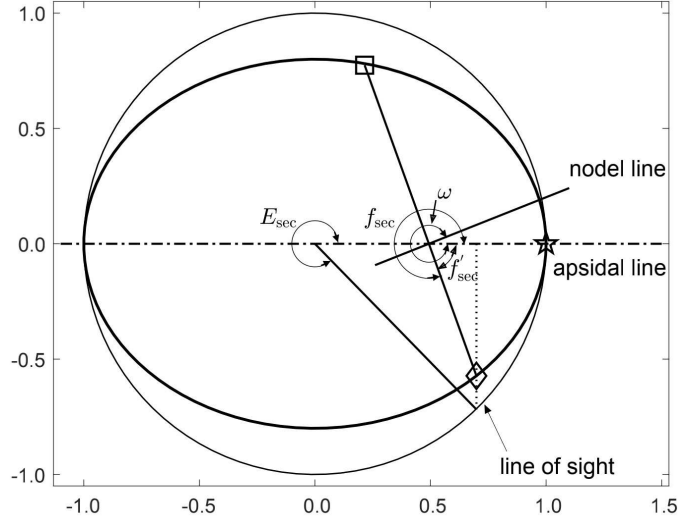}
  \caption{A diagram of an eclipse geometry as seen from above the elliptical orbit. The argument of periastron $\omega$ is measured from the ascending node crossing the plane of the sky to the periastron position at right (marked by the pentastar). The true anomaly $f_{\rm sec}$ at the secondary eclipse (diamond) is indicated as the angle from periastron to the stellar position. The angle from periastron to the position on the auxiliary circle is the eccentric anomaly $E$. The nodal line is perpendicular to the line of sight, and the true anomaly has a simple relationship $f_{\rm sec} + f'_{\rm sec} \equiv 2\pi$.}
  \label{Fig4}
\end{figure}

In this section, we demonstrate how to use the orbital phase of TEOs ($\phi_{\rm TEO}$) to determine the rate of apsidal motion arose from dynamic tides. It is well known that the geometrical change of the binary orbit causes the shift of pulsation phases. This section is an inverse derivation of this idea. We convert the delayed phase $\varphi_{\rm D}$ of TEOs to the tiny shift $\Delta \phi_{\rm TEO}$ of the orbital phase (Equation \ref{equ_add1}), and here derive back the changes of the periastron position due to tidal effects. Note that the derivation in this section is all about orbital phases.

Figure \ref{Fig4} illustrates the geometry of the elliptical orbit for an eccentric binary system \citep{gim1983, mat2016}. The components orbit the center of mass and attain the periastron (marked by the pentastar) at the right hand side of the diagram. Supposing the line of sight that we observe the binary is from the lower right, then the primary eclipse (marked by the square) occurs when the primary star is behind the secondary star along the line of sight, and the secondary eclipse occurs at the location marked by the diamond.  

The relation between the geometrical configuration and orbital phase is given by the \textit{Kepler} equation:

\begin{equation}
E_{\rm sec}\ - e\ {\rm sin}\ E_{\rm sec} = \frac{2\pi (t_{\rm sec}-t_{\rm p})}{P_{\rm a}} = 2\pi (\phi_{\rm sec} - \phi_{_{\rm TEO}})\ , \label{equ6}
\end{equation}
where $E_{\rm sec}$ is the eccentric anomaly of the secondary eclipse. As mentioned in Section \ref{sec3.2}, $\phi_{_{\rm TEO}} = t_{\rm p}/P_{\rm a}$ is the periastron passing time in units of orbital phase, and the tiny shift $\Delta \phi_{\rm TEO}$ of periastron position arising from the dynamic tides has been measured by delayed phase $\varphi_{\rm D}$ of TEOs  in Section \ref{sec3.2}. The phase difference between $\phi_{\rm sec}$ and $\phi_{_{\rm TEO}}$ indicates the time elapsed from the secondary eclipse to the position of tidally excited oscillations. The true anomaly at the secondary eclipse $f_{\rm sec}$ that measures the angle from periastron to the secondary eclipse can be derived from the relationship,

\begin{equation}
{\rm tan}\ \frac{f_{\rm sec}}{2} = \sqrt{\frac{1+e}{1-e}}\ {\rm tan}\ \frac{E_{\rm sec}}{2} \ . \label{equ7}
\end{equation}

The nodal line and the line of sight are orthogonal to each other. From inspection of Figure \ref{Fig4}, the argument of periastron is, 

\begin{equation}
\omega = f'_{\rm sec} + \frac{3\pi}{2} \ , \label{equ8}
\end{equation}
where the angle $f'_{\rm sec} \equiv  2\pi-f_{\rm sec}$. 

Solving the \textit{Kepler} equation, i.e. Equation (\ref{equ6}), and combining Equations (\ref{equ7})-(\ref{equ8}) thus yield the shifts of periastron argument caused by dynamic tides. The displacements of periastron are shown in red dots of Figure \ref{Fig3}. The slope of periastron argument  that is deduced from orbital phase shifts of dynamic tides is smaller than the slope of the total periastron movements. Specifically, the rate of apsidal motion is $\dot{\omega}_{_{\rm DT}} = 19.05\pm 1.70$ mrad yr$^{-1}$.

\section{Discussion and Conclusions}\label{sec5}

In previous studies, the lack of high-precision consecutive photometric data makes a barrier to probing  the impacts of dynamic tides on the apsidal motion from an observational point of view. Observations lasting for a few years are precious for the investigating of tidally excited oscillations. It allows us to measure the dynamic tides contribution to apsidal motion by tracing the delayed phase of TEOs. In this paper, we use the approximate one year long time baseline data from the NASA \textit{Kepler} space telescope to demonstrate that the dynamic tides contribution to apsidal motion is definitely present in an eccentric binary system KIC 4544587. 

Based on the high-precision \textit{Kepler} photometric data and the accurate absolute dimensions of  KIC 4544587, we infer the contributions of Newtonian term $\dot{\omega}_{\rm NT}$, general relativistic correction $\dot{\omega}_{\rm GR}$, and dynamic tides $\dot{\omega}_{_{\rm DT}}$ to the total rate of apsidal motion are $21.49 \pm 2.80$, $2.40 \pm 0.06$, and $19.05\pm 1.70$ mrad yr$^{-1}$, respectively. By considering the luminosity fluctuation with respect to periastron, we are accounting for the rotation of the orbit due to the Newtonian and general relativistic contribution. The sum of these three factors of $42.94 \pm 4.56$ mrad yr$^{-1}$ is in excellent agreement with the total rate of apsidal motion $42.97\pm 0.18$ mrad yr$^{-1}$ measured by mid-eclipse times. Furthermore, the consistency suggests that the Newtonian term is up to its upper limit, i.e. implies that the stellar rotation axis is aligned with the normal of the orbital plane in this eccentric binary system with tidally excited oscillations.

Comparing the different terms in the apsidal moment of this system, we find that the contribution of Newtonian term plays a dominant role in total apsidal motion, accounting for about 50\%, while the tidal effect accounts for about 44\% and the general relativistic correction for about 6\%.

The Newtonian term itself is the sum of effects induced by the density distribution of stars and the tidal deformation. Due to the lack of high-quality observational data in previous studies, the effects of dynamic tides were attributed to the Newtonian contribution. In other words, tidally excited oscillations were considered as factors of stellar internal structure and used to deduce the stellar internal structure constants $k_2$ \citep{sir2009}. However, when the tidal force of a companion star is large enough to excite the stellar oscillation modes in component stars, the contribution of dynamic tides will be comparable to that of Newtonian terms. In this scenario, the tidal effects can no longer be attributed to the stellar internal structure, but should be considered carefully. In previous literature, the theoretical rate of apsidal motion for some targets deviated from the observed rate \citep{gim2007}, perhaps because the contribution of dynamic tides was not considered sufficiently. The analysis method mentioned in this paper presents an alternative approaches to measure the contribution of dynamic tides quantitatively.

~\\
We acknowledge the anonymous referee for his/her suggestions that greatly improved this paper. O.J.W. gratefully acknowledges professor Ji-Lin Zhou (周济林) for teaching celestial mechanics knowledge. This work is supported by the National Key R\&D Program of China (2020YFC2201200), the National Natural Science Foundation of China (11803012), the science research grants from the China Manned Space Project (No. CMS-CSST-2021-B09), the Fundamental Research Funds for the Central Universities: Sun Yat-sen University (20lgpy174) and Nanjing University, and the Youth Science and Technology Talents Development Project of Guizhou Education Department (KY2018421). Y.C. has been supported by the National Natural Science Foundation of China (11521303, 11733010, and 11873103), Yunnan National Science Foundation (2014HB048), and Yunnan Province (2017HC018). J.C. is supported by a grant from the Max Planck Society to prepare for the scientific exploitation of the PLATO mission.


\end{CJK}
\end{document}